\documentstyle[preprint,pra,aps,eqsecnum,amsfonts,floats]{revtex}

\newtheorem{theorem}{Theorem}

\begin{document}

\draft

\title{Kinematic Orbits and the Structure of the Internal Space for
Systems of Five or More Bodies}

\author{Kevin A. Mitchell\thanks{KAMitchell@lbl.gov} and Robert
G. Littlejohn\thanks{RGLittlejohn@lbl.gov}}

\address{Department of Physics, University of California, 
Berkeley, California 94720}

\date{September 30, 1999}
\maketitle

\vskip 350pt

Short Title:  Kinematic Orbits and Systems of Five or More Bodies

PACS number(s): 31.15.-p,02.40.-k,02.20.-a

\newpage

\begin{abstract}
The internal space for a molecule, atom, or other $n$-body system can
be conveniently parameterised by $3n-9$ kinematic angles and three
kinematic invariants.  For a fixed set of kinematic invariants, the
kinematic angles parameterise a subspace, called a kinematic orbit,
of the $n$-body internal space.  Building on an earlier analysis of the three-
and four-body problems, we derive the form of these kinematic orbits
(that is, their topology) for the general $n$-body problem.  The case
$n=5$ is studied in detail, along with the previously studied cases
$n=3,4$.
\end{abstract}

\section{Introduction}

The group of kinematic rotations, called here the kinematic group, is
an important set of symmetries for the $n$-body kinetic energy.  In
fact, the kinematic group, to be defined precisely below, is the
largest (compact, connected) group of such symmetries acting on the
$n$-body internal space.  Not surprisingly then, the orbits (see
Appendix A, Ref.~\cite{Littlejohn98b}) of the kinematic group provide
a useful decomposition of the internal space.  It is the purpose of
this paper to analyse these orbits and to determine their topology.
We have been motivated by molecular applications, but the results are
quite general and could be applied to any $n$-body system with
rotational invariance, such as atoms or nuclei.

Although many reasons exist to study kinematic rotations and
their orbits, our current motivation derives from an interest in body
frame singularities and their implications for the quantum dynamics of
$n$-body systems.  In two previous papers
\cite{Littlejohn98b,Littlejohn98a}, body frame singularities in the
three- and four-body problems were studied explicitly.  The definition
of body frame singularities, their inevitability, the flexibility one
has in moving them, and their importance for quantum dynamics are all
discussed in Refs.~\cite{Littlejohn98b,Littlejohn98a}.  A detailed
study of frame singularities has also been made by Pack\cite{Pack94}.
Our earlier analysis of body frame singularities (especially the
singularities of the principal axis and related frames) involved
extensive use of kinematic rotations.  The present paper extends the
analysis of kinematic rotations to arbitrary $n$ and provides the
basis for a future discussion of frame singularities for the general
$n$-body problem.  Although the present paper concentrates on
kinematic orbits in their own right, for motivational reasons, we
provide a brief two paragraph account of frame singularities, referring
to Refs.~\cite{Littlejohn98b,Littlejohn98a,Pack94} for greater detail.

An early and necessary step in many quantum $n$-body computations is
choosing a set of body-fixed axes, that is, a body frame.  The
principal axis frame, in which the body-fixed axes are aligned with
the principal axes, is one common choice.  The body frame is a
function of the shape of the system, by which we mean the positions of
the bodies relative to each other; the shape may be parameterised by
$3n-6$ internal coordinates.  As has been previously noted
\cite{Littlejohn98b,Littlejohn98a,Pack94}, a body frame may fail to be
a smooth function of shape, and thus there may be points in the
internal, or shape, space at which it is singular.  (In this paper,
the internal space and shape space are synonymous.)  For example, in
the three-body problem, the principal axis frame is singular at all
oblate symmetric tops (among other shapes), and in the four-body
problem, the principal axis frame is singular at all symmetric tops
(among other shapes).  Body frame singularities have important
consequences for the form of the quantum wave function: roughly
speaking, the wave function has singularities matching those of the
body frame.  An understanding of body frame singularities is therefore
critical for understanding the singularities of the $n$-body wave
function.

The location of the body frame singularities in shape space depends on
the choice of body frame; by choosing different frames one can move these
singularities about or possibly remove them altogether (as is
essentially the case for the three-body problem
\cite{Littlejohn98a,Pack94}).  Thus, for many problems one can choose a frame
whose singularities are outside the physically relevant region of
shape space.  This is true of small vibration problems (about a
noncollinear equilibrium) in which the wave function is localised
around an equilibrium shape.  However, for scattering states and
delocalised bound states, it becomes harder to eliminate the
singularities from the region of interest.  For certain regions, it
becomes topologically impossible to remove them completely and one
must then understand their effects.

Though the study of kinematic orbits is developed here with the
ultimate intent of developing a deeper understanding of frame
singularities, several other reasons motivate our work.  First, since
the kinematic group is the largest (compact, connected) group of
symmetries (of the kinetic energy) acting on shape space, the
kinematic orbits provide an important foliation of shape space with
which to study the kinetic energy operator.  Furthermore, this
foliation suggests a convenient method of defining internal, or shape,
coordinates
\cite{Smith62,Zickendraht67,Zickendraht69,Zickendraht71,Keating85,Chapuisat91b,Chapuisat92,Aquilanti97}:
three internal coordinates are chosen to be kinematic invariants (for
example, the three principal moments of inertia), which label a
particular kinematic orbit, and the remaining $3n-9$ internal
coordinates are chosen to be kinematic angles, which parameterise the
position along the kinematic orbit.  Defining these angles and
properly specifying their ranges requires a clear understanding of the
topology of the orbits.  Finally, certain large amplitude internal
motions, such as pseudorotations, can be approximated by kinematic
rotations.  For such systems, it may be convenient to restrict the
region of physical interest to a single kinematic orbit.

The kinematic group is commonly viewed as the set of discrete
transformations between different conventions for Jacobi vectors.
Here, however, we define the kinematic group to be a continuous
symmetry group, namely $SO(n-1)$, which contains these
transformations.  The elements of the kinematic group $SO(n-1)$ are
called kinematic rotations to distinguish them from the ordinary
external $SO(3)$ rotations.  (Sometimes the terminology ``democracy
transformations'' and ``democracy group'' is used.)  Kinematic rotations
act (in the active sense) on the Jacobi vectors as shown in
Eq.~(\ref{r68}), from which one sees that they commute with external
rotations (as shown in Eq.~(\ref{r69})).  Therefore, kinematic
rotations do indeed have a well-defined action on the shape of an
$n$-body system.  It is the orbits of this action of the kinematic
rotations, for arbitrary $n$, which we compute in the present paper.

As examples of our general analysis, we specialise to the three- and
four-body problems, recovering the previously known results found in
Refs.~\cite{Littlejohn98b,Littlejohn98a,Littlejohn95}.  The kinematic
orbits for the three- and four-body cases do not exhibit the full
range of diversity found in the general $n$-body problem and are thus
somewhat special.  For example, in the four-body problem the kinematic
orbits can be classified by whether a shape is an asymmetric top, a
symmetric top, or a spherical top, a classification which does not
hold in the general $n$-body case.  We therefore also specialise to
the five-body problem, for which the results are not previously known.
As with the three- and four-body problems, the kinematic orbits for
the five-body problem have particularly simple forms.  However, there
are seven classes of kinematic orbits, which is representative of the
general $n$-body case.

The approach and methods used in this paper are geometrical in
nature.  We assume familiarity with the techniques of
Refs.~\cite{Littlejohn98b,Littlejohn98a} and some basic understanding
of Lie groups, their actions on manifolds, and the quotients by such
actions.  Appendix A of Ref.~\cite{Littlejohn98b} provides a useful
review, as do many basic texts \cite{Nakahara90,Bredon72}.

The structure of the paper is as follows.  Section~\ref{s2} contains
the principal derivations, in which we determine the isotropy subgroup
of the kinematic action on shape space.  This subgroup is related to
the kinematic orbit via Eq.~(\ref{r67}).  The results of
Sect.~\ref{s2} are summarised in Table~\ref{t1}.  The results for
arbitrary $n$ are discussed briefly in Sect.~\ref{s7} where we focus
primarily on the collinear shapes.  In Sect.~\ref{s3} we
specialise the results of Sect.~\ref{s2} to the three- and four-body
problems, and these results are summarised in Tables~\ref{t3} and
\ref{t2}.  Similarly, in Sect.~\ref{s4}, we specialise to the
five-body problem.  This requires substantially more work than the
three- and four-body cases which causes Sect.~\ref{s4} to constitute
almost half of the paper.  The five-body results are summarised in
Table~\ref{t4}.  Our conclusions are in Sect.~\ref{s5}.  We also
include an Appendix containing three important theorems on the actions
of Lie groups.

\section{The Topology of Kinematic Orbits for Arbitrary \mbox{\lowercase{$n$}}}

\label{s2}

In the centre of mass frame, the configuration of an $n$-body system
is parameterised by $n-1$ (mass-weighted) Jacobi vectors ${\bf r}_{s
\alpha}$, $\alpha = 1,..,n-1$.  Here, the $s$ subscript indicates
that the components of ${\bf r}_{s \alpha}$ are referred to a
space-fixed frame.  Jacobi vectors are a standard topic and we refer
to the literature for more details on their definition and
analysis\cite{Aquilanti86b,Littlejohn97}.  For notational convenience
we also introduce the $3 \times (n-1)$ matrix ${\sf F}_s$ whose
columns are the Jacobi vectors.  Explicitly, $F_{si
\alpha} = r_{s \alpha i}$, $i=1,2,3$, $\alpha = 1,..,n-1$, where $F_{si \alpha}$ and $r_{s \alpha i}$ are the components of ${\sf F}_s$ and ${\bf r}_{s \alpha}$ respectively.  

An ordinary rotation ${\sf Q} \in SO(3)$ acts on the Jacobi vectors by
standard multiplication on the left

\begin{eqnarray}
{\bf r}_{s \alpha} & \mapsto & {\sf Q} {\bf r}_{s \alpha}, 
\label{r69} \\
{\sf F}_s & \mapsto & {\sf Q} {\sf F}_s.
\end{eqnarray}
We call such a rotation an external rotation to distinguish it from a
kinematic rotation.  A kinematic rotation ${\sf K} \in SO(n-1)$ acts
by mixing up the $\alpha$ indices of the Jacobi vectors ${\bf r}_{s
\alpha}$,

\begin{eqnarray}
{\bf r}_{s \alpha} & \mapsto & \sum_\beta K_{\alpha \beta} {\bf r}_{s \beta}, 
\label{r68} \\
{\sf F}_s & \mapsto & {\sf F}_s{\sf K}^T,  
\end{eqnarray} 
where $K_{\alpha \beta}$ denotes the components of ${\sf K}$ and $T$
denotes the matrix transpose.  Notice that kinematic rotations commute
with all external rotations.  Thus, the kinematic group has a
well-defined action on the quotient of configuration space ${\Bbb
R}^{3n-3}$ by the group $SO(3)$ of external rotations.  This quotient
is called shape space, or the internal space, and its elements are called shapes.  A shape
thus determines the relative positions of the bodies with respect to
each other.  The purpose of this section is to find the topology of
the orbits of the kinematic group acting on shape space.

Considering a specific configuration ${\sf F}_s$ with shape $q$, the
kinematic orbit $\Gamma$ through $q$ is given by (that is, diffeomorphic to)

\begin{equation}
\Gamma = {SO(n-1) \over S}, 
\label{r67}
\end{equation}
where $S \subset SO(n-1)$ is the isotropy subgroup of the kinematic
action at $q$.  See Theorem~\ref{tm3} in the Appendix.  The isotropy
subgroup $S$ consists of all ${\sf K} \in SO(n-1)$ which leave the
shape $q$ invariant.  Our objective therefore is to determine $S$ for
each shape $q$.  Now, ${\sf F}_s{\sf K}^T$ and ${\sf F}_s$ have the
same shape if and only if they are related by a rotation ${\sf Q} \in SO(3)$.
Thus, our objective is to find all ${\sf K} \in SO(n-1)$ such that
there exists a ${\sf Q} \in SO(3)$ satisfying

\begin{equation}
{\sf F}_s{\sf K}^T = {\sf Q}{\sf F}_s.
\label{r10}
\end{equation}

We use the principal value decomposition to factor ${\sf F}_s$ into

\begin{equation}
{\sf F}_s = {\sf R} {\sf \Lambda} {\sf H}^T,
\label{r70}
\end{equation}
where ${\sf R} \in SO(3)$ and ${\sf H}\in SO(n-1)$, and ${\sf
\Lambda}$ is a $3 \times (n-1)$ matrix of the form

\begin{equation}
{\sf \Lambda} = 
\left[
\begin{array}{ccc|}
\lambda_1 & 0 & 0 \\
0 & \lambda_2 & 0 \\
0 & 0 & \lambda_3 \\
\end{array}
\right.
\overbrace{
\left.
\begin{array}{cccc}
0 & 0 & 0 & ... \\
0 & 0 & 0 & ... \\
0 & 0 & 0 & ... \\
\end{array}
\right].\!\!\!\!\!}^{n-4}
\end{equation}
Due to the nonuniqueness of the principal value decomposition, there
is no loss of generality in assuming that $\lambda_1, \lambda_2 \ge 0$
and that the $\lambda_i$'s are ordered as

\begin{equation}
\lambda_1 \ge \lambda_2
\ge |\lambda_3|.  
\label{r12}
\end{equation}
Using Eq.~(\ref{r70}), we recast the problem of
satisfying Eq.~(\ref{r10}) into the following problem: for which ${\sf
K} \in SO(n-1)$ does there exist a ${\sf Q} \in SO(3)$ such that

\begin{equation}
{\sf Q} {\sf \Lambda} ({\sf H}^T {\sf K} {\sf H})^T = {\sf \Lambda}. 
\end{equation}
The above equation shows that the isotropy subgroup $S$ of the action
of the kinematic group on the shape $q$ is conjugate to the isotropy
subgroup of the action of the kinematic group on the shape of the
configuration ${\sf \Lambda}$.  Replacing $S$ by a conjugate subgroup
in Eq.~(\ref{r67}) does not effect the resulting manifold $\Gamma$ (up to
diffeomorphism).  Thus, we assume without loss of generality that
${\sf F}_s = {\sf \Lambda}$, and hence we look to find all ${\sf K}
\in SO(n-1)$ such that there exists a ${\sf Q} \in SO(3)$ satisfying

\begin{equation}
{\sf Q} {\sf \Lambda} {\sf K}^T = {\sf \Lambda}. \label{r1}
\end{equation}
The answer depends on the rank of ${\sf \Lambda}$, which we denote by
$d$.  The quantity $d$ physically represents the dimensionality of the
shape q.  Thus, the $n$-body collision has $d=0$, linear shapes have
$d=1$, planar shapes have $d=2$, and full three-dimensional shapes
have $d=3$.  

For all values of $d$, ${\sf \Lambda}$ may be ``block diagonalised''
in the following manner

\begin{equation}
{\sf \Lambda} =
\left[
\begin{array}{c|c}
\parbox{35pt}{\vskip 14pt \hskip 12pt ${\sf \Sigma}$ \hskip 25pt \vskip 16pt } 
& \parbox{60pt}{\hskip 27pt {\sf 0} \hskip 26pt}  \\
\hline
{\sf 0} & \hskip .5 cm {\sf 0} \hskip .5cm
\end{array}
\right],
\end{equation}
where ${\sf \Sigma}$ is a $d\times d$ diagonal matrix and ${\sf 0}$
represents the zero matrices of the appropriate dimensions.  Equation~(\ref{r1}) can only be satisfied if ${\sf Q}$ and ${\sf K}$
are also block-diagonal, having the forms

\begin{equation}
{\sf K} 
= \left[
\begin{array}{c|c}
{\sf A} & {\sf 0} \\ \hline
{\sf 0} & {\sf B}
\end{array}
\right], 
\label{r2}
\end{equation}

\begin{equation}
{\sf Q} = 
\left[
\begin{array}{c|c}
{\sf C} & {\sf 0} \\ \hline
{\sf 0} & {\sf D} 
\end{array}
\right],
\end{equation}
where ${\sf C}$ and ${\sf A}$ are $d\times d$ matrices, ${\sf D}$ is a
$(3-d)\times (3-d)$ matrix, and ${\sf B}$ is an $(n-1-d)\times
(n-1-d)$ matrix.  Finding (special) orthogonal matrices ${\sf K}$ and
${\sf Q}$ satisfying Eq.~(\ref{r1}) is then equivalent to finding
orthogonal matrices ${\sf A}$,${\sf B}$,${\sf C}$,${\sf D}$ satisfying

\begin{eqnarray}
{\sf C}{\sf \Sigma}{\sf A}^T & = & {\sf \Sigma}, \label{r6} \\
\det {\sf A} \det {\sf B} & = & 1, \label{r8} \\
\det {\sf C} \det {\sf D} & = & 1. \label{r9}
\end{eqnarray}
The first equation follows from Eq.~(\ref{r1}).  The last two
equations ensure that ${\sf K}$ and ${\sf Q}$ have positive
determinant.

It can be shown, due to the fact that ${\sf \Sigma}$ is invertible and
${\sf A}, {\sf C} \in O(d)$, that Eq.~(\ref{r6}) can only be solved if
${\sf C} = {\sf A}$.  This in turn allows Eq.~(\ref{r9}) to be
rewritten as

\begin{equation}
\det {\sf A} \det {\sf D} = 1.
\label{r11}
\end{equation}
Recall that ${\sf D}$ is of dimension $3-d$.  Therefore, if $d = 3$,
the matrix ${\sf D}$ is completely eliminated from consideration, and
Eq.~(\ref{r11}) becomes simply 

\begin{equation}
\det {\sf A} = 1.
\end{equation}
However, if $d < 3$, then there  always exists an orthogonal matrix
${\sf D}$ such that $\det {\sf D} =
\det {\sf A}$.  Thus, Eq.~(\ref{r11}) imposes no constraint whatsoever on
${\sf A}$.  

For the sake of clarity, we now summarise the problem at hand.  For an
arbitrary diagonal $d \times d$ matrix ${\sf \Sigma}$, with nonzero eigenvalues
$\lambda_i$, $i=1,..,d$ satisfying Eq.~(\ref{r12}), we seek all matrices
${\sf K} \in SO(n-1)$ given by Eq.~(\ref{r2}), where the orthogonal $d
\times d$ matrix ${\sf A}$ and orthogonal $(n-1-d) \times (n-1-d)$
matrix ${\sf B}$ satisfy 

\begin{eqnarray}
{\sf A}{\sf \Sigma}{\sf A}^T 
& = & {\sf \Sigma}, 
\label{r3} \\
\det {\sf A} \det {\sf B} 
& = & 1, 
\label{r4} \\
\det {\sf A} 
& = & 1. \hskip 1cm \mbox{(required for $d = 3$ only)} 
\label{r5}
\end{eqnarray}
Notice that we have eliminated all reference to the matrix ${\sf Q}$.  

To proceed we consider each of the values of $d = 0,1,2,3$ separately.

\subsection*{Case $d = 3$}

From Eqs.~(\ref{r4}) and (\ref{r5}), we see that $\det{\sf A} =
\det{\sf B} = 1$.  Thus, ${\sf B}$ is an  element of $SO(n-4)$ and is independent of ${\sf A}$.  To find the allowed values of ${\sf A}$, we consider
three subcases: (i) all of the $\lambda_i$'s are distinct, (ii) two of
the $\lambda_i$'s are equal, the third is distinct, (iii) all of the
$\lambda_i$'s are equal.  Physically, these subcases correspond to
shapes which are asymmetric tops, symmetric tops, and spherical tops
respectively.

Assume subcase (i).  This is perhaps the most important class of
shapes since three-dimensional asymmetric tops are generic in shape
space.  From Eq.~(\ref{r3}) and the fact that ${\sf \Sigma}$ is
diagonal, ${\sf A}$ must be one of the four matrices in the group
$V_4$, where

\begin{equation}
V_4 
= \left\{
\left[
\begin{array}{ccc}
\;1\; & 0 & 0 \\
0 & \;1\; & 0 \\
0 & 0 & \;1\; \\
\end{array}
\right],
\left[
\begin{array}{ccc}
\;1\; & 0 & 0 \\
0 & -1 & 0 \\
0 & 0 & -1 \\
\end{array}
\right],
\left[
\begin{array}{ccc}
-1 & 0 & 0 \\
0 & \;1\; & 0 \\
0 & 0 & -1 \\
\end{array}
\right],
\left[
\begin{array}{ccc}
-1 & 0 & 0 \\
0 & -1 & 0 \\
0 & 0 & \;1\; \\
\end{array}
\right]
\right\}.
\label{r13}
\end{equation}
The group $V_4$ is called the viergruppe.  It played a critical roll
in earlier analysis of the four-body problem
\cite{Littlejohn98b,Littlejohn95}; we will reproduce part of this
earlier analysis in Sect.~\ref{s3}.  Thus, ${\sf K}$ lives in a
subgroup of $SO(n-1)$ isomorphic to $V_4\times SO(n-4)$.  The group
$V_4\times SO(n-4)$ is therefore the isotropy subgroup $S$ for
three-dimensional asymmetric tops.  

Assume subcase (ii). Since ${\sf \Sigma}$ is diagonal with two equal
eigenvalues, ${\sf A}$ must be block-diagonal, with a $2\times 2$
block which can be any element ${\sf S} \in O(2)$ and a $1 \times 1$
block which must be $\det {\sf S}$ to ensure the condition $\det {\sf
A} = 1$.  Thus, ${\sf A}$ lives in a subset of $SO(3)$ isomorphic to
$O(2)$ and hence $O(2) \times SO(n-4)$ is the isotropy subgroup $S$
for three-dimensional symmetric tops.

Assume subcase (iii).  Since all eigenvalues of ${\sf \Sigma}$ are
equal, ${\sf \Sigma}$ is proportional to the identity.  Hence, ${\sf
A}$ can be any element of $SO(3)$, and hence $SO(3) \times SO(n-4)$ is
the isotropy subgroup $S$ for three-dimensional spherical tops.

\subsection*{Case $d=2$}

We consider two subcases: (i) $\lambda_1 \ne \lambda_2$ (ii)
$\lambda_1 = \lambda_2$.  Physically, these correspond to asymmetric
and symmetric tops respectively.

Assume subcase (i).  Since ${\sf \Sigma}$ is diagonal with two
different eigenvalues, ${\sf A}$ must be one of the four matrices in
the group $V_4$, where

\begin{equation}
V_4 
= \left\{
\left[
\begin{array}{cc}
\;1\; & 0 \\
0 & \;1\; \\
\end{array}
\right],
\left[
\begin{array}{cc}
\;1\; & 0  \\
0 & -1 \\
\end{array}
\right],
\left[
\begin{array}{cc}
-1 & 0 \\
0 & \;1\;  \\
\end{array}
\right],
\left[
\begin{array}{cc}
-1 & 0 \\
0 & -1 \\
\end{array}
\right]
\right\}.
\label{r48}
\end{equation}
This is another representation of the viergruppe in Eq.~(\ref{r13}).
For notational simplicity, we use the same symbol for both groups; it
will be clear from context which group is intended.  Since ${\sf B}
\in O(n-3)$, the matrix ${\sf K}$ is in $V_4 \times O(n-3)$.  We still
must apply the one remaining constraint given by Eq.~(\ref{r4}).  For this reason, we
introduce the notation $V_4
\times_{\det +1} O(n-3)$ for all elements in  $V_4 \times
O(n-3)$ with unit determinant.  Thus, $V_4
\times_{\det +1} O(n-3)$ is the isotropy subgroup $S$ for two-dimensional asymmetric tops.

Assume subcase (ii).  Since ${\sf \Sigma}$ is diagonal with two equal
eigenvalues, ${\sf \Sigma}$ is proportional to the identity.  Thus the
matrix ${\sf A}$ can be any element of $O(2)$, and hence $O(2)
\times_{\det +1} O(n-3)$ is the isotropy subgroup $S$ for planar
symmetric tops.

\subsection*{Case $d = 1$}

Since ${\sf A}$ is in $O(1)$, ${\sf A}$ is either $1$ or $-1$.  From
Eq.~(\ref{r4}), ${\sf A} = \det {\sf B}$.  Therefore, ${\sf B} \in O(n-2)$
completely determines ${\sf K}$.  Thus, $O(n-2)$ is the the isotropy
subgroup $S$ for linear shapes.

\subsection*{Case $d = 0$}
Since the dimension of ${\sf A}$ is $0$ here, there is really no
matrix ${\sf A}$ to worry about.  That is, ${\sf K} = {\sf B}$.  Thus,
$SO(n-1)$ is the isotropy subgroup $S$ for the $n$-body collision.

\section{Comments on the General Results}

\label{s7}
We summarise the results from the preceding section in Table~\ref{t1}
and comment on a few special cases.  First, as was to be expected the
kinematic orbit passing through the $n$-body collision is a single
point $\Gamma = SO(n-1)/SO(n-1) = \{0\}$.

A more interesting case is that of the collinear shapes.  It is a
well-known fact that 

\begin{equation}
{SO(k+1) \over O(k)} 
= {\Bbb R}P^{k}, 
\label{r71}
\end{equation}
where ${\Bbb R}P^k$ is the $k$-dimensional real projective space ($k
\ge 1$).  The $k$-dimensional real projective space is the space of
lines in ${\Bbb R}^{k+1}$.  It may also be viewed as the
$k$-dimensional sphere $S^k$ with antipodal points identified.  A
quick proof of Eq.~(\ref{r71}) can be given with the aid of
Theorem~\ref{tm3}, taking

\begin{equation}
M = {\Bbb R}P^k 
= \left\{ \{ \hat{\bf e}, -\hat{\bf e} \} 
| \hat{\bf e} = (\hat{e}_1, ..., \hat{e}_{k+1}) \in S^k \subset {\Bbb R}^{k+1}\right\}
\end{equation} 
and $G = SO(k+1)$.  A matrix ${\sf K} \in SO(k+1)$ maps $\{ \hat{\bf
e}, -\hat{\bf e} \}$ into $\{ {\sf K} \hat{\bf e}, -{\sf K}\hat{\bf e}
\}$.  If $\hat{\bf e} = (1, 0, ..., 0)$, then one can see that the
isotropy subgroup of $\{ \hat{\bf e}, -\hat{\bf e} \}$ is $H = O(k)$.
(In fact, $H$ is exactly the same representation of $O(k)$, $k=n-2$,
discussed above for the case $d = 1$.)  Since the orbit of the action
of $SO(k+1)$ on ${\Bbb R}P^k$ is the entire space ${\Bbb R}P^k$,
Eq.~(\ref{r71}) follows from Theorem~\ref{tm1}.

Applying Eq.~(\ref{r71}), we see that the kinematic orbit of a
collinear shape is

\begin{equation}
\Gamma 
= {SO(n-1) \over O(n-2)} 
= {\Bbb R}P^{n-2}. 
\label{r72}
\end{equation}
For the four-body problem (in three-dimensions) the two-fragment exit
channels can be visualised as seven pairs of antipodal points on
$S^2$ or, equivalently, seven points on ${\Bbb R}P^2$
\cite{Littlejohn98b,Aquilanti96b}.  Kinematic angles between these 
points were explicitly computed in Ref.~\cite{Aquilanti96b}.  These
results were based on the understanding that ${\Bbb R}P^2$ is the
kinematic orbit for collinear shapes in the four-body
problem.\footnote{More precisely, the analysis of
Ref.~\cite{Aquilanti96b} is based on the fact that $S^2$ is the
kinematic orbit for shapes in the one-dimensional four-body problem.
We ignore the one-dimensional $n$-body problem in this paper.}  (The
interest in collinear shapes stems from the fact that a two-fragment
state becomes more and more collinear as the separation between the
fragments increases.)  The four-body work was an extension of
well-known results for the three-body problem (in three-dimensions) in
which the two-fragment exit channels can be visualised as three points
on a circle $S^1 = {\Bbb R}P^1$ with certain kinematic angles between
them.  The result presented in Eq.~(\ref{r72}) shows that in general
the two-fragment exit channels can be viewed as points on ${\Bbb
R}P^{n-2}$, or equivalently, pairs of antipodal points on $S^{n-2}$.
(Some quick combinatorics gives the number of points on ${\Bbb
R}P^{n-2}$ to be $2^{n-1} - 1$.)  Of course, this is only a
topological result, and we say nothing about the values of the
kinematic angles between such points.

\begin{table}[ht]
\caption{Isotropy subgroups of the kinematic action on shape space}
\label{t1}
\begin{tabular}{|c|@{\hspace{.5cm}}c@{\hspace{.5cm}}|
@{\hspace{.5cm}}c@{\hspace{.5cm}}|
@{\hspace{.5cm}}c@{\hspace{.5cm}}|}
Class & Physical Description of Class & Isotropy Subgroup $S$ 
& $\mbox{dim }(\Gamma)$\tablenote{$\Gamma = $ kinematic orbit $= SO(n-1)/S$} \\
\hline
3(i)  & 3D asymmetric top & $ V_4 \times SO(n-4) $ 
& $3n-9$ \\
3(ii) & 3D symmetric top  & $ O(2) \times SO(n-4) $ 
& $3n - 10$ \\
3(iii)& 3D spherical top  & $ SO(3) \times SO(n-4) $ 
& $3n - 12$ \\
2(i)  & Planar asymmetric top            & $ V_4 \times_{\det +1} O(n-3) $ 
& $2n - 5$ \\
2(ii) & Planar symmetric top             & $ O(2) \times_{\det +1} O(n-3) $
& $2n - 6$ \\
1     & Linear shape                           & $ O(n-2)$ 
& $n - 2$ \\
0     & $n$-body collision               & $ SO(n-1)$ 
& $0$  \\
\end{tabular} 
\end{table}

\section{The Three- and Four-Body Problems}

\label{s3}

We specialise the preceding analysis to the three- and four-body
problems.  These cases have been studied earlier
\cite{Littlejohn98b,Littlejohn98a,Littlejohn95}.  The present analysis
serves both as a check on the general results in Sect.~\ref{s2} and as
practice for the five-body problem.

We begin with the three-body problem $n=3$.  In the analysis of
Sect.~\ref{s2} we assumed for convenience that $n \ge 4$.  However, by
closely examining this analysis, one sees that the results presented
in Table~\ref{t1} are also valid for $n=3$, so long as one ignores the
nonsensical results for the three-dimensional classes 3(i), 3(ii), and
3(iii).  For the classes 2(i) and 2(ii), the factor $O(n-3) = O(0)$ of
$S$ is to be ignored.  Thus, the isotropy subgroup of the class 2(i) is
the two-element group $S = {\Bbb Z}_2$ consisting of those elements of
$V_4$ in Eq.~(\ref{r48}) with unit determinant.  Similarly, the isotropy subgroup of the
class 2(ii) is $SO(2)$.  For the class 1, the isotropy subgroup is $S = O(1)
= \{+1, -1\} = {\Bbb Z}_2$.  These results are summarised in
Table~\ref{t3}.  

\begin{table}[ht]
\caption{Isotropy subgroups and kinematic orbits for the three-body problem}
\label{t3}
\begin{tabular}{|c|@{\hspace{.5cm}}c@{\hspace{.5cm}}|
@{\hspace{.5cm}}c@{\hspace{.5cm}}|
@{\hspace{.5cm}}c@{\hspace{.5cm}}|
@{\hspace{.5cm}}c@{\hspace{.5cm}}|}
Class & Physical Description of Class & $S$
& $\mbox{dim }(\Gamma)$
& $\Gamma$ \tablenote{$\Gamma = \mbox{kinematic orbit} = SO(2)/S$} \\
\hline
2(i)  & Planar asymmetric top            & $ {\Bbb Z}_2 $
& $1$  & $S^1$ \\
2(ii) & Planar symmetric top             & $ SO(2) $
& $0$  & $\{0\}$ \\
1     & Linear shape                          & ${\Bbb Z}_2$
& $1$  & $S^1$ \\
0     & $3$-body collision               & $ SO(2)$
& $0$ & $\{0\}$ \\
\end{tabular}
\end{table}

Since the kinematic group $SO(2)$ is particularly simple, the topology
of the kinematic orbits may be presented in a more direct and
illuminating form than the quotient $SO(2)/S$.  These forms are
recorded in the rightmost column of Table~\ref{t3}.  For the
three-body problem, we can provide a convenient picture of the
kinematic rotations which explains why the kinematic orbits have the
topologies that they do.  The three-body shape space is conveniently
parameterised by three internal coordinates $(w_1, w_2, w_3)$ with
ranges $-\infty < w_1, w_2 < \infty$, $0 \le w_3 < \infty$.  The
kinematic rotations act on shape space via standard $SO(3)$ matrices
rotating about the $w_3$-axis.  It so happens that the $w_3$-axis
consists of the planar (noncollinear) symmetric tops as well as the
3-body collision.  Thus, the kinematic orbits of these shapes contain
a single point, whereas the kinematic orbits of all other shapes
are circles about the $w_3$-axis.

Turning to the four-body problem, we specialise the entries of
Table~\ref{t1} for $n=4$ and display the results in Table~\ref{t2}.
For the classes 3(i), 3(ii), and 3(iii) we ignore the factor $SO(n-4)
= SO(0)$.  For the classes 2(i) and 2(ii), the factor $O(n-3)$ reduces
to $O(1) = \{ +1, -1 \}$.  Since the choice of $+1$ or $-1$ in $O(1)$
is fixed by the $\det = +1$ constraint, the isotropy subgroup for classes
2(i) and 2(ii) are $V_4$ and $O(2)$ respectively.  An interesting
observation is that the results for the four-body problem are
classified solely on the basis of the symmetries of the moment of
inertia tensor.  That is, the topology of the kinematic orbit depends
only on whether a shape is a spherical top, symmetric top, or
asymmetric top.  (This fact is not true for $n=3$ or $n\ge 5$.)

\begin{table}[ht]
\caption{Isotropy subgroups and kinematic orbits for the four-body problem}
\label{t2}
\begin{tabular}{|c|@{\hspace{.5cm}}c@{\hspace{.5cm}}|
@{\hspace{.5cm}}c@{\hspace{.5cm}}|
@{\hspace{.5cm}}c@{\hspace{.5cm}}|
@{\hspace{.5cm}}c@{\hspace{.5cm}}|}
Class & Physical Description of Class & $S$
& $\mbox{dim }(\Gamma)$
& $\Gamma$ \tablenote{$\Gamma = \mbox{kinematic orbit} = SO(3)/S$} \\
\hline
3(i)  & 3D asymmetric top & $V_4$
& $3$ & $S^3/V_8$ \\
3(ii) & 3D symmetric top  & $O(2)$
& $2$ & ${\Bbb R}P^2$ \\
3(iii)& 3D spherical top  & $ SO(3) $
& $0$ & $\{0\}$ \\
2(i)  & Planar asymmetric top            & $ V_4 $
& $3$ & $S^3/V_8$ \\
2(ii) & Planar symmetric top             & $ O(2) $
& $2$ & ${\Bbb R}P^2$ \\
1     & Linear shape                          & $ O(2)$
& $2$ & ${\Bbb R}P^2$ \\
0     & $4$-body collision               & $ SO(3)$
& $0$ & $\{0\}$ \\
\end{tabular}
\end{table}

In the final column of Table~\ref{t2}, we have again represented the
topologies of the kinematic orbits in a more direct and illuminating
form than simply the quotient $SO(3)/S$.  We already explained in
Sect.~\ref{s7}, how the equality $SO(3)/SO(2) = {\Bbb R}P^2$ comes about.  Thus, the only 
orbit which requires special attention here is

\begin{equation}
\Gamma 
= {SO(3) \over V_4}
= {SU(2) \over V_8}
= {S^3 \over V_8}.
\label{r59}
\end{equation}
Here $V_8$ is an eight element
subgroup of SU(2)

\begin{equation}
V_8 = \{ \pm{\sf I}, \pm{\sf \omega}_1,\pm{\sf \omega}_2,\pm{\sf \omega}_3\},
\label{r28}
\end{equation} 
where $\omega_i = -i \sigma_i$, $i=1,2,3$, and the
$\sigma_i$'s are the usual Pauli matrices.  Explicitly,

\begin{eqnarray}
{\sf \omega}_1 & = & -i{\sf \sigma}_1 =
\left[
\begin{array}{cc}
0 & -i \\
-i & 0 \\
\end{array}
\right], 
\label{r61} \\
{\sf \omega}_2 
& = & -i{\sf \sigma}_2
= 
\left[
\begin{array}{cc}
0 & -1 \\
\; 1 \;& 0 \\
\end{array}
\right], 
\label{r62} \\
{\sf \omega}_3 
& = & -i{\sf \sigma}_3
= \left[
\begin{array}{cc}
-i\; & 0 \\
0 & \; i \;\\
\end{array}
\right]. 
\label{r63}
\end{eqnarray}
The matrices ${\sf \omega}_i$ satisfy

\begin{eqnarray}
{\sf \omega}_1 {\sf \omega}_2 
& = &  -{\sf \omega}_2 {\sf \omega}_1 = {\sf \omega}_3, 
\label{r42} \\
{\sf \omega}_2 {\sf \omega}_3 
& = &  -{\sf \omega}_3 {\sf \omega}_2 = {\sf \omega}_1, 
\label{r43} \\
{\sf \omega}_3 {\sf \omega}_1 
& = &  -{\sf \omega}_1 {\sf \omega}_3 = {\sf \omega}_2, 
\label{r44} \\
{\sf \omega}_i^\dagger
& = & {\sf \omega}^{-1}_i = - {\sf \omega}_i. 
\label{r45} 
\end{eqnarray}
These product rules show that $V_8$ is the quaternion group.  To prove
Eq.~(\ref{r59}) we recall some basic facts about $SO(3)$.  First,
$SU(2)$ is the double cover of $SO(3)$, and we denote the projection by
$\pi:SU(2) \rightarrow SO(3)$.  The kernel of $\pi$ is ${\Bbb Z}_2 =
\{ {\sf I}, -{\sf I} \}$ and hence $SO(3) = SU(2)/{\Bbb Z}_2$.
If ${\sf R} = \pi({\sf U})$ for some ${\sf U} \in SU(2)$,
then ${\sf R}$ can be given explicitly by

\begin{equation}
R_{ij} 
= - {1 \over 2} \mbox{tr}\;({\sf \omega}_i {\sf U} 
{\sf \omega}_j {\sf U}^\dagger),
\label{r37}
\end{equation}
where $R_{ij}$, $i,j = 1,2,3$, are the components of ${\sf R}$.  Using
the product rules Eqs.~(\ref{r42}) -- (\ref{r45}) and Eq.~(\ref{r37}),
one can verify that $\pi(V_8) = V_4$ given in Eq.~(\ref{r13}), and
hence $V_4 = V_8/{\Bbb Z}_2$.  We now employ Theorem~\ref{tm2} from
the Appendix, with $G = V_8$, $H = {\Bbb Z}_2 = \{ {\sf I}, -{\sf I}
\}$, $M = SU(2)$.  (It is trivial to verify that ${\Bbb Z}_2$ is
normal in $V_8$.)  Hence, Eq.~(\ref{r60}) yields

\begin{equation}
{SO(3) \over V_4}
= {SU(2)/{\Bbb Z}_2 \over V_8/{\Bbb Z}_2} 
= {SU(2) \over V_8}
= {S^3 \over V_8}, 
\end{equation}
where we recall that $SU(2)$ is diffeomorphic to the three-dimensional
sphere $S^3$.

\section{The Five-Body Problem}

\label{s4} 

The entries of Table~\ref{t1} are specialised for the five-body
problem, $n=5$, and displayed in Table~\ref{t4}.  As with the three-
and four-body cases, we also represent the topology of the kinematic
orbits in a more direct and illuminating form than the original
quotient $\Gamma = SO(4)/S$.    

\begin{table}[ht]
\caption{Isotropy subgroups and kinematic orbits for the five-body problem}
\label{t4}
\begin{tabular}{|c|@{\hspace{.5cm}}c@{\hspace{.5cm}}|
@{\hspace{.5cm}}c@{\hspace{.5cm}}|
@{\hspace{.5cm}}c@{\hspace{.5cm}}|
@{\hspace{.5cm}}c@{\hspace{.5cm}}|}
Class & Physical Description of Class & $S$ 
& $\mbox{dim }(\Gamma)$
& $\Gamma$ \tablenote{$\Gamma = \mbox{kinematic orbit} = SO(4)/S$} \\
\hline
3(i)  
& 3D asymmetric top 
& $ V_4 $ 
& $6$ & $S^3 \times (S^3/V_8)$ \\
3(ii) 
& 3D symmetric top  
& $ O(2) $ 
& $5$ & $S^3 \times {\Bbb R}P^2$ \\
3(iii)
& 3D spherical top  
& $ SO(3) $ 
& $3$ & $S^3$ \\
2(i)  
& Planar asymmetric top            
& $ V_4 \times_{\det +1} O(2) $ 
& $5$ & ${\Bbb R}P^3 \times {\Bbb R}P^2$ \\
2(ii) 
& Planar symmetric top             
& $ O(2) \times_{\det +1} O(2) $
& $4$ & $(S^2 \times S^2)/{\Bbb Z}_2$ \\
1     
& Linear shape                           
& $ O(3)$ 
& $3$ & ${\Bbb R}P^3$ \\
0     
& $5$-body collision               
& $ SO(4)$ 
& $0$ & $\{0\}$ \\
\end{tabular} 
\end{table}

Before deriving the results in Table~\ref{t4}, we make a few
observations.  First, for all classes but 2(ii) we present the
kinematic orbits as products of well-studied two- and
three-dimensional manifolds.  In fact, all of these simpler manifolds
already appear in the four-body problem, either as kinematic orbits or
as the group manifold $SO(3) = {\Bbb R}P^3$.  One obvious advantage of
such a simple description of the topologies of these spaces is that it
simplifies the introduction of kinematic angles or some other
parameterisation of the orbits.  For example, in the case of a 3D
asymmetric top, we may introduce six kinematic angles by taking three
to be standard Euler angles on $S^3 = SU(2)$ and three to be Euler
angles on $S^3/V_8$.  The ranges of the latter three angles must be
carefully restricted to account for the fact that $S^3/V_8$ is only
one eighth the size of $S^3$.  A discussion of the ranges of such
angles has already been given in the context of the four-body problem
\cite{Aquilanti97,Littlejohn95,Kuppermann94,Kuppermann97}.  Of
course, the space $S^3/V_8$ can be parameterised in various other
ways, such as the convenient coordinates $\bbox{\tau} =
(\tau_1,\tau_2,\tau_3)$ suggested by Reinsch
\cite{Littlejohn98b,Reinsch96}.  

It is interesting to note that the kinematic orbit of a collinear shape
is ${\Bbb R}P^3 = SO(3)$.  A point on such an orbit (such as one of
the two-fragment exit channels discussed in Sect.~\ref{s7}) can
therefore be identified with a rotation matrix and may in turn be
parameterised by one of the many standard parameterisations of $SO(3)$
(Euler angles, axis-angle variables, Cayley-Klein parameters, etc.).

We proceed now to derive the results of Table~\ref{t4}.

\subsection{The Projection $\pi$ from $SU(2)\times SU(2)$ to $SO(4)$}

With the four-body problem, we have found it useful to work with the double
cover $SU(2)$ of the kinematic group $SO(3)$.  With the five-body
problem, we also find it useful to work with the double cover of the
kinematic group.  In this case, the kinematic group is $SO(4)$ and its
double cover is $SU(2) \times SU(2)$.  In this section we give an
explicit realization of the projection from $SU(2)\times SU(2)$ to
$SO(4)$.

First, we introduce the function $g$ which maps a complex number into
a corresponding $2 \times 2$ real matrix.  Explicitly,

\begin{equation}
g(a + ib) = 
\left[
\begin{array}{cc}
\;a\; & -b \\
b & a \\
\end{array}
\right],
\end{equation}
where $a$ and $b$ are real numbers.  The function $g$ is real linear
and preserves multiplication.  Specifically, it is straightforward to
verify the following identities

\begin{eqnarray}
g(z_1 + az_2) & = & g(z_1) + a g(z_2), 
\label{r77} \\
g(z_1 z_2)    & = & g(z_1) g(z_2), \\
g( z_1^*)     & = & g(z_1)^T, \\
\mbox{tr}\; g(z_1) & = & z_1 + z_1^* = 2\mbox{Re}\; z_1, \\
\det g(z_1) & = & |z_1|^2, \\
\mbox{if } z_1 \ne 0 \mbox{ then} \hskip 1cm g(z_1^{-1})   & = & g(z_1)^{-1},
\label{r78}
\end{eqnarray}
where $z_1$ and $z_2$ are complex, $z_1^*$ is the complex conjugate of
$z_1$, and $a$ is real.  We define $g$ acting on a $k \times k$
complex matrix to be the $2k \times 2k$ real matrix given by

\begin{equation}
g\left(
\left[
\begin{array}{ccc}
a_{11} + i b_{11} & a_{12} + i b_{12} & \dots \\
a_{21} + i b_{21} & a_{22} + i b_{22} & \dots \\
\vdots            & \vdots            & \ddots \\
\end{array}
\right]\right)
=
\left[
\begin{array}{ccccc}
\;a_{11}\; & -b_{11} & \;a_{12}\;  & -b_{12} & \dots \\
b_{11} &  a_{11} & b_{12}  & a_{12} & \dots \\
a_{21} & -b_{21} & a_{22}  & -b_{22} & \dots \\
b_{21} &  a_{21} & b_{22}  & a_{22} & \dots \\
\vdots & \vdots  & \vdots  & \vdots & \ddots \\
\end{array}
\right].
\end{equation}
The following identities are analogous to Eqs.~(\ref{r77}) -- (\ref{r78}),

\begin{eqnarray}
g({\sf M} + a {\sf N}) & = & g({\sf M}) + a g({\sf N}), 
\label{r14} \\
g({\sf M}{\sf N}) & = & g({\sf M}) g({\sf N}), 
\label{r15} \\
g({\sf M}^\dagger) & = & g({\sf M})^T, 
\label{r16} \\
\mbox{tr}\; g({\sf M}) & = & \mbox{tr}\;{\sf M} + (\mbox{tr}\;{\sf M})^*
= 2 \mbox{Re}\; (\mbox{tr}\;{\sf M}), 
\label{r79} \\
\det g({\sf M}) & = & |\det {\sf M}|^2, 
\label{r75} \\
\mbox{if } {\sf M} \mbox{ invertible then} 
\hskip 1cm  g({\sf M}^{-1}) & = & g({\sf M})^{-1}, 
\label{r17}
\end{eqnarray}
where ${\sf M}$ and ${\sf N}$ are square complex matrices, ${\sf
M}^\dagger$ is the Hermitian conjugate of ${\sf M}$, and $a$ is a real
number.  Except for Eq.~(\ref{r75}), these identities are relatively
straightforward to prove.  To prove Eq.~(\ref{r75}), we first assume
that ${\sf M}$ is normal and invertible, which allows us to write
${\sf M} = \exp{\sf X}$ for some matrix ${\sf X}$.  Then,

\begin{eqnarray}
\det g({\sf M}) 
& = & \det [g(\exp{\sf X})] 
= \det [\exp g({\sf X})] 
= \exp [\mbox{tr}\; g({\sf X})] 
= \exp [\mbox{tr}\; {\sf X} + (\mbox{tr}\; {\sf X})^*] \nonumber \\
& = & \det (\exp {\sf X}) [\det (\exp {\sf X})]^*
= |\det {\sf M}|^2,
\end{eqnarray}
where we have used Eq.~(\ref{r79}) and Eq.~(\ref{r21}) (which appears
below) as well as the fact that $\det \exp = \exp \mbox{tr}$.  We now
consider an arbitrary (possibly non-normal) invertible matrix ${\sf
M}$ and note that it may be written as a product of normal matrices
(using, for example, polar or principal value decompositions).  Using
this fact and Eq.~(\ref{r15}), we observe that Eq.~(\ref{r75}) holds for
${\sf M}$ as well.  Having shown that Eq.~(\ref{r75}) is valid for all
invertible matrices, analytic continuation shows that it is valid for
all matrices.

We now define $\pi$ from $SU(2)\times SU(2)$ to $SO(4)$ by

\begin{equation}
\pi({\sf U}_1, {\sf U}_2)
= g({\sf U}_1) {\sf P} g({\sf U}_2) {\sf P}^T,
\label{r20}
\end{equation}
where

\begin{equation}
{\sf P} 
=
{1 \over \sqrt{2}}
\left[
\begin{array}{cccc}
1 & \;1\; & \;0\; & 0 \\
-1& 1 & 0 & 0 \\
0 & 0 & 1 & 1 \\
0 & 0 & 1 &-1 \\
\end{array}
\right] \in SO(4),
\end{equation}
and where ${\sf U}_1,{\sf U}_2 \in SU(2)$. From Eqs.~(\ref{r16}) --
(\ref{r17}), we observe that $g({\sf U}_1)$ and
$g({\sf U}_2)$ are in $SO(4)$.  Since ${\sf P}$ is also in $SO(4)$ we
verify that $\pi({\sf U}_1, {\sf U}_2)$ is in $SO(4)$.  To verify that
$\pi$ is a group homomorphism we must verify the following equation

\begin{equation}
\pi({\sf U}_1 {\sf V}_1^{-1}, {\sf U}_2 {\sf V}_2^{-1})
= \pi({\sf U}_1, {\sf U}_2)
\pi({\sf V}_1, {\sf V}_2)^{-1},
\label{r19}
\end{equation}
where ${\sf U}_1, {\sf U}_2, {\sf V}_1, {\sf V}_2 \in SU(2)$ are
arbitrary.  To prove Eq.~(\ref{r19}), we first hypothesise that 

\begin{equation}
g({\sf U}) \left[{\sf P} g({\sf V}) {\sf P}^T\right]
= \left[{\sf P} g({\sf V}) {\sf P}^T \right]g({\sf U}),
\label{r18}
\end{equation} 
where ${\sf U}, {\sf V} \in SU(2)$ are arbitrary.  We postpone the
proof of Eq.~(\ref{r18}) temporarily in order to show how it is used
to prove Eq.~(\ref{r19}).  To this end, we have

\begin{eqnarray}
\pi({\sf U}_1 {\sf V}_1^{-1}, {\sf U}_2 {\sf V}_2^{-1})
& = & g({\sf U}_1 {\sf V}_1^{-1}) 
{\sf P} g({\sf U}_2 {\sf V}_2^{-1}){\sf P}^T
= g({\sf U}_1)g({\sf V}_1)^{-1}
{\sf P} g({\sf U}_2)g({\sf V}_2)^{-1}{\sf P}^T \nonumber \\
& = &  g({\sf U}_1)g({\sf V}_1)^{-1}
\left[{\sf P} g({\sf U}_2) {\sf P}^T\right]
\left[{\sf P} g({\sf V}_2)^{-1}{\sf P}^T\right]
\nonumber \\
& = & \left[g({\sf U}_1) {\sf P} g({\sf U}_2) {\sf P}^T \right]
\left[{\sf P} g({\sf V}_2)^{-1}{\sf P}^Tg({\sf V}_1)^{-1}\right]
\nonumber \\
& = & \pi({\sf U}_1, {\sf U}_2)
\pi({\sf V}_1, {\sf V}_2)^{-1},
\end{eqnarray}
where the first equality follows from the definition Eq.~(\ref{r20}),
the second from Eqs.~(\ref{r15}) and (\ref{r17}), the third from
inserting ${\sf P}^T{\sf P} = {\sf I}$, the forth from Eq.~(\ref{r18}), and
the final equality again from Eq.~(\ref{r20}).
 
We return now to prove Eq.~(\ref{r18}).  We find it convenient to work
with the Lie algebras $su(2)$ and $so(4)$ of $SU(2)$ and $SO(4)$
respectively.  We see from Eq.~(\ref{r16}) that if ${\sf X} \in su(2)$,
that is ${\sf X}$ is a $2 \times 2$ anti-Hermitian matrix, then
$g({\sf X}) \in so(4)$, that is, $g({\sf X})$ is a $4 \times 4$
antisymmetric real matrix.  Furthermore,

\begin{eqnarray}
g(\exp{\sf X}) & = & \exp g({\sf X}), 
\label{r21} \\
{\sf P} \exp({\sf X}) {\sf P}^T 
& = & \exp( {\sf P}{\sf X} {\sf P}^T),
\label{r22}
\end{eqnarray}
where Eq.~(\ref{r21}) follows from Eqs.~(\ref{r14}) and (\ref{r15}).
Taking ${\sf U} = \exp{\sf X}, {\sf V} = \exp{\sf Y}$,
Eqs.~(\ref{r21}) and (\ref{r22}) allow Eq.~(\ref{r18}) to be
reexpressed as

\begin{equation}
\exp [g({\sf X})] \exp [ {\sf P} g({\sf Y}) {\sf P}^T]
=  \exp [{\sf P} g({\sf Y}) {\sf P}^T] \exp [g({\sf X})].
\label{r24}
\end{equation} 
The above equation is valid so long as 

\begin{equation}
[g({\sf X}),{\sf P} g({\sf Y})
{\sf P}^T] 
= 0
\label{r73}
\end{equation}
for arbitrary ${\sf X}, {\sf Y} \in su(2)$, where $[\; \; , \; \; ]$
is the matrix commutator.

We prove Eq.~(\ref{r73}) by using a basis of $su(2)$, which we choose
to be the matrices $\omega_i$, $i=1,2,3$, given in Eqs.~(\ref{r61}) --
(\ref{r63}).  From the product rules Eqs.~(\ref{r42}) -- (\ref{r44}),
this basis satisfies the Lie algebra relations

\begin{equation}
[{\sf \omega}_i, {\sf \omega}_j] 
= 2 \sum_k \epsilon_{ijk} {\sf \omega}_k.
\end{equation}
It is straightforward to compute the matrices ${\sf J}_i = g({\sf
\omega}_i)$ and ${\sf L}_i = {\sf P} g({\sf
\omega}_i) {\sf P}^T$,

\begin{eqnarray}
{\sf J}_1 
& = & g({\sf \omega}_1) 
=
\left[
\begin{array}{cccc}
0 & \;0\; & 0 & \;1\; \\
0 & 0 &-1 & 0 \\
0 & 1 & 0 & 0 \\
-1& 0 & 0 & 0 \\
\end{array}
\right],
\label{r26} \\
{\sf J}_2 
& = & g({\sf \omega}_2) 
=
\left[
\begin{array}{cccc}
\;0\; & \;0\; &-1 & 0 \\
0 & 0 & 0 &-1 \\
1 & 0 & 0 & 0 \\
0 & 1 & 0 & 0 \\
\end{array}
\right], 
\label{r33} \\
{\sf J}_3 
& = & g({\sf \omega}_3) 
=
\left[
\begin{array}{cccc}
0 & \;1\; & \;0\; & 0 \\
-1& 0 & 0 & 0 \\
0 & 0 & 0 &-1 \\
0 & 0 & 1 & 0 \\
\end{array}
\right], 
\label{r34} \\
{\sf L}_1 
& = & {\sf P} g({\sf \omega}_1){\sf P}^T 
=
\left[
\begin{array}{cccc}
\;0\; & \;0\; & 0 &-1 \\
0 & 0 &-1 & 0 \\
0 & 1 & 0 & 0 \\
1 & 0 & 0 & 0 \\
\end{array}
\right], 
\label{r32} \\
{\sf L}_2 
& = & {\sf P} g({\sf \omega}_2){\sf P}^T 
=
\left[
\begin{array}{cccc}
\;0\; & 0 &-1 & \;0\; \\
0 & 0 & 0 & 1 \\
1 & 0 & 0 & 0 \\
0 &-1 & 0 & 0 \\
\end{array}
\right], 
\label{r35} \\
{\sf L}_3 
& = & {\sf P} g({\sf \omega}_3){\sf P}^T 
=
\left[
\begin{array}{cccc}
0 & \;1\; & 0 & \;0\; \\
-1& 0 & 0 & 0 \\
0 & 0 & 0 & 1 \\
0 & 0 &-1 & 0 \\
\end{array}
\right].
\label{r27}
\end{eqnarray}
The matrices ${\sf J}_i$, $i = 1,2,3$ and ${\sf L}_i$, $i = 1,2,3$ are
a basis of the Lie algebra $so(4)$ and it is straightforward to show
that they satisfy the Lie algebra relations

\begin{eqnarray}
[{\sf J}_i, {\sf J}_j] 
& = & 2 \sum_k \epsilon_{ijk} {\sf J}_k, \\
\; [{\sf L}_i, {\sf L}_j] 
& = & 2 \sum_k \epsilon_{ijk} {\sf L}_k, \\
\; [{\sf J}_i, {\sf L}_j ] & = & 0.
\label{r23}
\end{eqnarray}
The above equations exhibit the well-known fact that $so(4) = su(2)
\oplus su(2)$.  Since the matrices ${\sf J}_i$ and ${\sf L}_i$ span
the space of matrices of the form $g({\sf X})$ and ${\sf P}g({\sf
Y}){\sf P}^T$ respectively (${\sf X}, {\sf Y} \in su(2)$),
Eq.~(\ref{r23}) proves Eq.~(\ref{r73}), from which follows
Eqs.~(\ref{r24}), (\ref{r18}), and (\ref{r19}).  We have thus shown
$\pi$ to be a group homomorphism.

The mapping $\pi$ has several important properties which we will use
later.  First, it follows directly from the definition Eq.~(\ref{r20})
that for arbitrary group elements $({\sf U}_1, {\sf U}_2)
\in SU(2) \times SU(2)$

\begin{eqnarray}
\pi({\sf U}_1, -{\sf U}_2) 
& = & \pi(-{\sf U}_1, {\sf U}_2) 
= -\pi({\sf U}_1, {\sf U}_2), 
\label{r49} \\
\pi(-{\sf U}_1, -{\sf U}_2) 
& = & \pi({\sf U}_1, {\sf U}_2). 
\label{r25}
\end{eqnarray}
Second, in light of Eq.~(\ref{r23}), the definition of $\pi$ can be
conveniently reexpressed using the Lie algebra, 

\begin{equation}
\pi(\exp{\sf X}_1, \exp{\sf X}_2)
= \exp [ g({\sf X}_1) + {\sf P} g({\sf X}_2) {\sf P}^T].
\label{r31}
\end{equation}
Since the matrices ${\sf J}_i$ and ${\sf L}_i$ form a basis of the Lie
algebra $so(4)$, the above equation shows that $\pi$ is surjective.
However, $\pi$ is obviously not injective since  $\pi({\sf I},
{\sf I})= \pi(-{\sf I}, -{\sf I}) = {\sf I}$.  In fact, $({\sf I}, {\sf
I})$ and $(-{\sf I}, -{\sf I})$ are the only two elements of $SU(2)
\times SU(2)$ which map to ${\sf I} \in SO(4)$.  To prove this 
 we consider two arbitrary elements ${\sf U}_1, {\sf U}_2 \in SU(2)$,
 expressed in axis-angle form as

\begin{eqnarray}
{\sf U}_1 
& = & \cos \theta_1 {\sf I} - \sin \theta_1 \hat{\bf n}_1 \cdot \bbox{\omega}, \\
{\sf U}_2 
& = & \cos \theta_2 {\sf I} - \sin \theta_2 \hat{\bf n}_2 \cdot \bbox{\omega}, 
\end{eqnarray}
where $\theta_1$, $\theta_2$ are rotation angles, $\hat{\bf n}_1$,
$\hat{\bf n}_2$ are rotation axes, and $\bbox{\omega} = ({\sf
\omega}_1, {\sf \omega}_2, {\sf
\omega}_3)$.  Then,

\begin{eqnarray}
\pi({\sf U}_1, {\sf U}_2) 
& = & (\cos \theta_1 {\sf I} - \sin \theta_1 \hat{\bf n}_1 \cdot {\bf
J}) (\cos \theta_2 {\sf I} - \sin \theta_2 \hat{\bf n}_2 \cdot {\bf
L})
\nonumber \\
& = & \cos \theta_1 \cos \theta_2 {\sf I}
- \sin \theta_1 \cos \theta_2 (\hat{\bf n}_1 \cdot {\bf J}) \nonumber \\
&& - \cos \theta_1 \sin \theta_2 (\hat{\bf n}_2 \cdot {\bf L})
+ \sin \theta_1 \sin \theta_2 
(\hat{\bf n}_1 \cdot {\bf J})(\hat{\bf n}_2 \cdot {\bf L}),
\end{eqnarray}
where ${\bf J} = ({\sf J}_1, {\sf J}_2, {\sf J}_3)$ and ${\bf L} =
({\sf L}_1, {\sf L}_2, {\sf L}_3)$.  It can easily be verified that
$\{{\sf I}, {\sf J}_i, {\sf L}_i, {\sf J}_i{\sf L}_j\}$ forms a basis
of all $4\times 4$ real matrices.  Thus, if $\pi({\sf U}_1, {\sf U}_2)
= {\sf I}$ then 

\begin{eqnarray}
\cos \theta_1 \cos \theta_2 & = & 1, \\
\sin \theta_1 \cos \theta_2 & = & 0, \\
\cos \theta_1 \sin \theta_2 & = & 0, \\
\sin \theta_1 \sin \theta_2 & = & 0, 
\end{eqnarray}
which only occurs if $\theta_1 = \theta_2 = 0$ or $\theta_1 = \theta_2
= \pi$, corresponding to ${\sf U}_1 = {\sf U}_2 = {\sf I}$ or ${\sf
U}_1 = {\sf U}_2 = -{\sf I}$ respectively.

In summary, we have proved that $\pi$ given by Eq.~(\ref{r20}) is a two-to-one
surjective group homomorphism from $SU(2)\times SU(2)$ to $SO(4)$.

\subsection{The double covers of the isotropy subgroups}

For each isotropy subgroup $S$, we determine its double cover $\hat{S} =
\pi^{-1}(S)$.  That is we must find the two elements of $SU(2) \times
SU(2)$ which map to each element of $S$.  In light of Eq.~(\ref{r25}),
these two elements are related by a minus sign, that is, $({\sf U}_1,
{\sf U}_2) \in \hat{S}$ and $(-{\sf U}_1, -{\sf U}_2) \in
\hat{S}$ map to the same element in $S$.  Thus,  the
problem of determining $\hat{S}$ reduces to finding only one element in
$SU(2)\times SU(2)$ which maps to each element in $S$.  

Since ${\Bbb Z}_2 = \{ ({\sf I}, {\sf I}), -({\sf I}, {\sf I}) \}$ is
obviously normal in $\hat{S}$ and $\bbox{(}SU(2) \times SU(2)\bbox{)}/{\Bbb Z}_2 =
SO(4)$ and $\hat{S} /{\Bbb Z}_2 = S$, we apply Theorem~\ref{tm2} to find

\begin{equation}
\Gamma 
= {SO(4) \over S} 
= {\bbox{(}SU(2) \times SU(2)\bbox{)}/{\Bbb Z}_2 \over \hat{S} / {\Bbb Z}_2}
= {SU(2) \times SU(2)\over \hat{S} }.
\label{r74}
\end{equation}
We will use this result extensively to determine the topology of the
kinematic orbits.  We analyse each class in Table~\ref{t4} separately.

\subsubsection{The class 3(i) of 3D asymmetric tops}

Considering the analysis of Sect.~\ref{s2}, an element ${\sf K}$ of
$S$ depends on the two matrices ${\sf A}$ and ${\sf B}$ as shown in
Eq.~(\ref{r2}).  Considering the class 3(i) for $n=5$, the matrix
${\sf B}$ is simply the $1 \times 1$ matrix ${\sf B} = 1$.  The matrix
${\sf A}$ must belong to the group $V_4$ given by Eq.~(\ref{r13}).
Thus, the group $S$ contains the following matrices

\begin{eqnarray}
S & = & \{ {\sf I}, {\sf E}_1, {\sf E}_2, {\sf E}_3 \}, \\
{\sf E}_1 & = & 
\left[
\begin{array}{cccc}
\;1\; & 0 & 0 & \;0\; \\
0 & -1 & 0 & 0 \\
0 & 0 & -1 & 0 \\
0 & 0 & 0  & 1 \\
\end{array}
\right], \\
{\sf E}_2 & = & 
\left[
\begin{array}{cccc}
-1 & \;0\; & 0 & \;0\; \\
0 & 1 & 0 & 0 \\
0 & 0 & -1 & 0 \\
0 & 0 & 0  & 1 \\
\end{array}
\right], \\
{\sf E}_3 & = & 
\left[
\begin{array}{cccc}
-1 & 0 & \;0\; & \;0\; \\
0 & -1 & 0 & 0 \\
0 & 0  & 1 & 0 \\
0 & 0  & 0 & 1 \\
\end{array}
\right].
\end{eqnarray}
From Eq.~(\ref{r20}) we see that $\pi(\omega_i,\omega_i) = {\sf J}_i
{\sf L}_i$, where ${\sf J}_i$ and ${\sf L}_i$ are given by
Eqs.~(\ref{r26}) -- (\ref{r27}). By direct matrix multiplication, we
find

\begin{equation}
\pi({\sf \omega}_i, {\sf \omega}_i) 
= {\sf J}_i {\sf L}_i 
= {\sf E}_i.
\label{r30}
\end{equation}
Thus, having found one element in $SU(2) \times SU(2)$ which maps to
each element of $S$, the double cover $\hat{S}$ is the eight element
group

\begin{equation}
\hat{S} = \left\{ \pm({\sf I}, {\sf I}), 
\pm({\sf \omega}_1, {\sf \omega}_1),
\pm({\sf \omega}_2, {\sf \omega}_2),
\pm({\sf \omega}_3, {\sf \omega}_3)
\right \} = \tilde{V}_8.
\end{equation}
In the above, the tilde over $V_8$ has a technical meaning which we
now define.  If $H$ is an arbitrary subgroup of $SU(2)$, then
$\tilde{H}$ is an isomorphic subgroup of $SU(2) \times SU(2)$ as shown
in Eq.~(\ref{r29}).  Thus, $\hat{S}$ is isomorphic to the quaternion
group in Eq.~(\ref{r28}).  Since $\hat{S}$ has the form of $\tilde{H}$
in Eq.~(\ref{r29}), we apply Theorem~\ref{tm1} from the Appendix to
find

\begin{equation}
\Gamma = {SU(2) \times SU(2) \over \hat{S}}
= SU(2) \times {SU(2) \over V_8} 
= S^3 \times {S^3 \over V_8}.
\end{equation}

\subsubsection{The class 3(ii) of 3D symmetric tops}

For the class 3(ii), ${\sf B}$ is again the $1 \times 1$ matrix ${\sf
B} = 1$.  The matrix ${\sf A}$ consists of a $2 \times 2$ block ${\sf
S} \in O(2)$ and a $1 \times 1$ block $\det {\sf S}$.  Combining these
results into the single matrix ${\sf K}$, we see that $S$ consists of

\begin{equation}
S = \left \{
\begin{array}{c|c}
\left[
\begin{array}{c|c}
\parbox{35pt}{\vskip 16pt \hskip 14pt ${\sf S}$ \hskip 25pt \vskip 18pt } 
& {\sf 0} \\ \hline
{\sf 0} & 
\begin{array}{cc}
\det {\sf S} & 0 \\
0            & 1 
\end{array}
\end{array}
\right]
&
{\sf S} \in O(2) 
\end{array}
\right \}
= 
\left \{ {\sf I}, {\sf E}_1 \right \}
\left \{ 
\begin{array}{c|c}
\left[
\begin{array}{c|c}
\;{\sf S}\; & \;{\sf 0}\; \\ \hline
{\sf 0} & {\sf I}
\end{array}
\right]
& {\sf S} \in SO(2)
\end{array}
\right \},
\label{r36}
\end{equation} 
where in the second equality we have factored $S$ into the product of
two groups.  Having encountered the elements of the first factor
earlier, we recall that $\pi({\sf I}, {\sf I}) = {\sf I}$ and (from
Eq.~(\ref{r30})) that $\pi({\sf \omega}_1, {\sf \omega}_1) = {\sf
E}_1$.  Considering the second factor, we note

\begin{eqnarray}
\pi\bbox{(}\exp(\theta {\sf \omega}_3), \exp(\theta {\sf \omega}_3)\bbox{)}
& = & \exp\bbox{(} \theta({\sf J}_3 + {\sf L}_3)\bbox{)}
= \exp \left( 
2 \theta
\left[
\begin{array}{c|c}
\begin{array}{cc}
0 & \;1\; \\
-1 & 0 
\end{array}
& {\sf 0} \\ \hline
{\sf 0} & 
\parbox{35pt}{\vskip 16pt \hskip 15pt ${\sf 0}$ \hskip 25pt \vskip 18pt } 
\end{array}
\right]
\right) \nonumber \\
& = & 
\left[
\begin{array}{c|c}
\begin{array}{cc}
\cos 2\theta & \sin 2\theta \\
- \sin 2\theta  & \cos 2\theta 
\end{array}
& {\sf 0} \\ \hline
{\sf 0} & 
\parbox{35pt}{\vskip 16pt \hskip 16pt ${\sf I}$ \hskip 25pt \vskip 18pt } 
\end{array}
\right],
\label{r39}
\end{eqnarray}
where we used Eqs.~(\ref{r31}), (\ref{r34}), and (\ref{r27}).
Equation~(\ref{r39}) shows that the double cover of the second factor in
Eq.~(\ref{r36}) is the group

\begin{equation}
\tilde{A} 
= \left \{ 
({\sf U}, {\sf U}) | {\sf U} \in A \right \},
\label{r50}
\end{equation}
where $A$ is the group

\begin{equation}
A 
= \{ \exp (\theta {\sf \omega}_3)
| 0 \le \theta < 2 \pi
\}.
\label{r76}
\end{equation}
Note that $A = U(1) \subset SU(2)$.  The double cover of $S$ is
therefore

\begin{equation}
\hat{S} 
= \left \{ 
({\sf I}, {\sf I}),
({\sf \omega}_1, {\sf \omega}_1 ) 
\right \}
\tilde{A}
= \left \{
({\sf U}, {\sf U}) | {\sf U} \in B \right \} = \tilde{B},
\label{r52}
\end{equation}
where $B$ is the group

\begin{equation}
B 
= \{ {\sf I}, {\sf \omega}_1 \} A.
\label{r41}
\end{equation}

A priori, it is perhaps not obvious that $B$ is actually a group.
To verify that $B$ is indeed closed under multiplication and
inverses, the following identity is useful

\begin{equation}
{\sf \omega}_1 \exp(\theta {\sf \omega}_3) {\sf \omega}_1^\dagger
= \exp(\theta {\sf \omega}_1 {\sf \omega}_3 {\sf \omega}_1^\dagger)
= \exp(- \theta {\sf \omega}_3),
\label{r40}
\end{equation}
which derives from Eqs.~(\ref{r42}) -- (\ref{r45}).  When forming products
and inverses of the elements of $B$, Eq.~(\ref{r40}) allows any
${\sf \omega}_1$ factors to be shifted to the left so that the final
result again has the form displayed in Eq.~(\ref{r41}).

Since $\hat{S}$ has the form of Eq.~(\ref{r29}), we apply  Theorem~\ref{tm1}
to find

\begin{equation}
\Gamma 
= {SU(2) \times SU(2) \over \hat{S}}
= {SU(2) \times SU(2) \over \tilde{B}}
= SU(2) \times {SU(2) \over B}. 
\label{r46}
\end{equation}
The quotient $SU(2)/B$ is diffeomorphic to ${\Bbb R} P^2$.  To prove
this, we first consider the quotient $SU(2)/A = SU(2)/U(1)$.  It is
well known that $SU(2)/U(1) = S^2$.  One way of seeing this fact is to
consider the action of $SU(2)$ on ${\Bbb R}^3$ via the $3 \times 3$
orthogonal matrices given in Eq.~(\ref{r37}).  The orbit of $SU(2)$
acting on $\hat{\bf z}\in {\Bbb R}^3$ is clearly the sphere $S^2$.
The isotropy subgroup of the vector $\hat{\bf z}$ is the group $A$ of
$U(1)$ rotations about the $\hat{\bf z}$-axis.  Theorem~\ref{tm3} thus
gives the desired result

\begin{equation}
{SU(2) \over A} = S^2.  
\label{r64}
\end{equation}
Furthermore, we note the following explicit identification between a
right coset $\bbox{[}{\sf U}\bbox{]} = A {\sf U} \in SU(2)/A$ and a
unit vector $\hat{\bf n} \in S^2$ (denoting a coset with bold square
brackets),

\begin{equation}
\bbox{[} {\sf U} \bbox{]} \leftrightarrow \hat{\bf n}
= {\sf R}^T \hat{\bf z} 
= -{1 \over 2} \mbox{tr}\;(\bbox{\omega} {\sf U}^\dagger {\sf \omega}_3 {\sf U}),
\label{r38}
\end{equation}
where ${\sf R}$ is given by Eq.~(\ref{r37}) and
$\bbox{\omega} = ({\sf \omega}_1, {\sf \omega}_2, {\sf \omega}_3)$.
We have placed the transpose on ${\sf R}$ in order that $\hat{\bf n}$
be well-defined for right cosets; observe that the right hand side of
Eq.~(\ref{r38}) is invariant under ${\sf U}
\mapsto \exp(\theta {\sf \omega}_3) {\sf U}$.

Having computed $SU(2)/A$, we apply Theorem~\ref{tm2} to compute
$SU(2)/B$.  With regards to the notation of the theorem, we take
$G=B$, $H=A$ and $M=SU(2)$. We first must verify that $A$ is normal in
$B$. Proving this fact reduces to showing that ${\sf
\omega}_1 \exp(\theta {\sf \omega}_3) {\sf \omega}_1^\dagger$ is in $A$
for an arbitrary $\exp(\theta {\sf \omega}_3)\in A$.  This fact, in
turn, follows immediately from Eq.~(\ref{r40}).  Thus, $A$ is normal
in $B$ and $B / A$ is a well-defined group isomorphic to ${\Bbb
Z}_2$.  According to Theorem~\ref{tm2}, the non-identity element
$\bbox{[}{\sf \omega}_1\bbox{]} \in B/A$ acts on $\bbox{[}{\sf U}\bbox{]} \in SU(2)/A$ by
$\bbox{[}{\sf \omega}_1\bbox{]}\bbox{[}{\sf U}\bbox{]} = \bbox{[}{\sf \omega}_1 {\sf U}\bbox{]}$.  Identifying
$\bbox{[}{\sf U}\bbox{]}$ with $\hat{\bf n}$, the action of $\bbox{[}{\sf \omega}_1\bbox{]}$ on
$\hat{\bf n}$ is

\begin{equation} 
\bbox{[}{\sf \omega}_1\bbox{]} \hat{\bf n}
= -{1 \over 2} 
\mbox{tr}\;
[\bbox{\omega} ({\sf \omega}_1{\sf U})^\dagger 
 {\sf \omega}_3 ({\sf \omega}_1 {\sf U})] 
= {1 \over 2} 
\mbox{tr}\;(\bbox{\omega} {\sf U}^\dagger  {\sf \omega}_3  {\sf U}) 
= -\hat{\bf n},
\label{r65}
\end{equation}
which follows from Eqs.~(\ref{r42}) -- (\ref{r45}).
Thus, the quotient $S^2/{\Bbb Z}_2$ is ${\Bbb R}P^2$, and by
Theorem~\ref{tm2} we have the following identifications

\begin{equation}
{SU(2) \over B} 
= {SU(2)/A \over B/A}
= {S^2 \over  {\Bbb Z}_2}
= {\Bbb R}P^2.
\label{r55}
\end{equation}
Recalling Eq.~(\ref{r46}), we find that 

\begin{equation}
\Gamma = S^3 \times {\Bbb R}P^2.
\end{equation}

\subsubsection{The class 3(iii) of 3D spherical  tops}

For the class 3(iii), ${\sf B}$ is again the $1 \times 1$ matrix ${\sf
B} = 1$.  The matrix ${\sf A}$ can be any matrix in $SO(3)$.  Thus,
the group $S$ is

\begin{equation}
S = \left \{
\begin{array}{c|c}
\left[
\begin{array}{c|c}
\parbox{35pt}{\vskip 16pt \hskip 13pt ${\sf A}$ \hskip 25pt \vskip 18pt } 
& {\sf 0} \\ \hline
\;{\sf 0}\; & 1
\end{array}
\right]
&
{\sf A} \in SO(3) 
\end{array}
\right \}.
\label{r57}
\end{equation}
To find the double cover of $S$, we consider an arbitrary matrix ${\sf
U} \in SU(2)$ expressed as ${\sf U} = \exp\left({\bf n} \cdot
\bbox{\omega}\right)$ for some vector ${\bf n} = (n_1, n_2, n_3)$.  Then
we find from Eq.~(\ref{r31}) and Eqs.~(\ref{r26}) -- (\ref{r27}) that

\begin{equation}
\pi({\sf U}, {\sf U}) 
= \pi\bbox{(}\exp\left( {\bf n} \cdot \bbox{\omega}\right), 
\exp\left({\bf n} \cdot \bbox{\omega}\right)\bbox{)} 
= \exp\left[ {\bf n} \cdot ({\bf J} + {\bf L} )\right].  
\label{r47}
\end{equation}
Furthermore, from Eqs.~(\ref{r26}) -- (\ref{r27}), we see that

\begin{eqnarray}
{\sf J}_1 + {\sf L}_1 
& = & 2
\left[
\begin{array}{cccc}
\;0\; & \;0\; & 0 & \;0\; \\
0 & 0 &-1 & 0 \\
0 & 1 & 0 & 0 \\
0 & 0 & 0 & 0 \\
\end{array}
\right], \\
{\sf J}_2 + {\sf L}_2 
& = & 2
\left[
\begin{array}{cccc}
\;0\; & \;0\; &-1 & \;0\; \\
0 & 0 & 0 & 0 \\
1 & 0 & 0 & 0 \\
0 & 0 & 0 & 0 \\
\end{array}
\right], \\
{\sf J}_3 + {\sf L}_3 
& = & 2
\left[
\begin{array}{cccc}
0 & \;1\; & \;0\; & \;0\; \\
-1& 0 & 0 & 0 \\
0 & 0 & 0 & 0 \\
0 & 0 & 0 & 0 \\
\end{array}
\right].
\end{eqnarray}
Thus, the matrices ${\sf J}_i + {\sf L}_i$ generate $S$, and from
Eq.~(\ref{r47}) the double cover of $S$ is

\begin{equation}
\hat{S} = \{ ({\sf U}, {\sf U}) | {\sf U} \in SU(2) \} = \tilde{E}.
\label{r58}
\end{equation}
Applying Theorem~\ref{tm1}, we have

\begin{equation}
\Gamma = {SU(2) \times SU(2) \over \hat{S}} 
= {SU(2) \times SU(2) \over \tilde{E}} 
= SU(2) \times {SU(2) \over SU(2)} = S^3.
\label{r66}
\end{equation}

\subsubsection{The class 2(i) of planar asymmetric tops}

For the class 2(i), the matrices ${\sf A}$ and ${\sf B}$ are
respectively in $V_4$ (as shown in Eq.~(\ref{r48})) and $O(2)$.  These
matrices must further satisfy $\det {\sf A} {\sf B} = 1$. For
convenience, we switch the positions of ${\sf A}$ and ${\sf B}$ in
Eq.~(\ref{r2}).  That is, we place ${\sf A}$ in the lower right block
and ${\sf B}$ in the upper left block.  This switch is equivalent to
conjugating by a permutation and hence does not effect the topology of
the quotient $SO(4)/S$.  With this modification, the isotropy subgroup
is

\begin{eqnarray}
S & = & \left \{
\begin{array}{c|c}
\left[
\begin{array}{c|c}
\;{\sf B}\; & \;{\sf 0}\; \\ \hline
{\sf 0} & {\sf A}
\end{array}
\right]
&
{\sf B} \in O(2), {\sf A} \in V_4, \det {\sf A} {\sf B} = 1 
\end{array}
\right \} \nonumber \\
& = &
\left \{ {\sf I}, {\sf E}_1, -{\sf E}_2, -{\sf E}_3 \right \}
\left \{ 
\begin{array}{c|c}
\left[
\begin{array}{c|c}
\;{\sf B}\; & \;{\sf 0}\; \\ \hline
{\sf 0} & {\sf I}
\end{array}
\right]
& {\sf B} \in SO(2)
\end{array}
\right \},
\label{r51}
\end{eqnarray} 
where we have again factored $S$ into the product of two groups.
Considering the elements in the first factor, we may combine
Eqs.~(\ref{r49}) and (\ref{r30}) to produce

\begin{eqnarray}
\pi({\sf \omega}_1, {\sf \omega}_1) 
& = & {\sf E}_1, \\
\pi({\sf \omega}_2, -{\sf \omega}_2) 
& = & -{\sf E}_2, \\
\pi(-{\sf \omega}_3, {\sf \omega}_3) 
& = & -{\sf E}_3. 
\end{eqnarray}
The second factor of Eq.~(\ref{r51}) is identical to the second factor
of Eq.~(\ref{r36}), and hence the double cover of the second factor is
$\tilde{A}$ given by Eq.~(\ref{r50}).  Thus, the double cover of $S$
is

\begin{eqnarray}
\hat{S} 
& = & \left \{ 
({\sf I}, {\sf I}),
({\sf \omega}_1, {\sf \omega}_1 ),
({\sf \omega}_2,-{\sf \omega}_2 ),
(-{\sf \omega}_3,{\sf \omega}_3 )
\right \}
\tilde{A} \nonumber \\
& = & 
\{ 
({\sf I}, {\sf I}),
({\sf \omega}_2, -{\sf \omega}_2 )
\}
\{ 
({\sf I}, {\sf I}),
({\sf \omega}_1, {\sf \omega}_1 )
\}
\tilde{A}
=
\{ 
({\sf I}, {\sf I}),
({\sf \omega}_2, -{\sf \omega}_2 )
\}
\tilde{B},
\end{eqnarray}
where $\tilde{B}$ is given in Eq.~(\ref{r52}).

We apply Theorem~\ref{tm2} to determine the topology of the kinematic
orbit, taking $G = \hat{S}$, $H = \tilde{B}$, and $M = SU(2) \times
SU(2)$.  Using Eqs.~(\ref{r42}) -- (\ref{r45}), it is straightforward
to verify that $\tilde{B}$ is normal in $\hat{S}$, and hence
$\hat{S}/\tilde{B}$ is a well-defined group isomorphic to ${\Bbb
Z}_2$.  The action of the nonidentity element $\bbox{[}{\sf \omega}_2, -{\sf
\omega}_2\bbox{]} \in \hat{S}/\tilde{B}$ on $\bbox{[}{\sf U}_1, {\sf U}_2\bbox{]} \in
\bbox{(}SU(2) \times SU(2)\bbox{)}/\tilde{B}$ is

\begin{equation}
\bbox{[}{\sf \omega}_2, -{\sf \omega}_2\bbox{]}\bbox{[}{\sf U}_1, {\sf U}_2\bbox{]}
= \bbox{[}{\sf \omega}_2{\sf U}_1, -{\sf \omega}_2{\sf U}_2\bbox{]}.  
\end{equation}
Using Theorem~\ref{tm1}, we previously showed that $\bbox{(}SU(2)
\times SU(2)\bbox{)}/\tilde{B}$ is diffeomorphic to $SU(2) \times
\bbox{(}SU(2)/B\bbox{)}$.  The diffeomorphism is given by
Eq.~(\ref{r53}).  Using this diffeomorphism we find that the action of
$\bbox{[}{\sf \omega}_2, -{\sf \omega}_2\bbox{]} \in
\hat{S}/\tilde{B}$ on $({\sf U}_1, \bbox{[}{\sf U}_2\bbox{]}) \in
SU(2) \times \bbox{(}SU(2)/B\bbox{)}$ is

\begin{equation}
\bbox{[}{\sf \omega}_2, -{\sf \omega}_2\bbox{]}({\sf U}_1, \bbox{[}{\sf U}_2\bbox{]})
= ( -{\sf U}_1, \bbox{[}-{\sf \omega}_2{\sf U}_2\bbox{]})  
= ( -{\sf U}_1, \bbox{[}{\sf U}_2\bbox{]}),  
\label{r54}
\end{equation}
where the last equality follows from the fact that $-{\sf \omega}_2 =
{\sf \omega}_1 {\sf \omega}_3 \in B$.  (Be careful not to confuse the
bold square bracket notation $\bbox{[} \; \; , \; \; \bbox{]}$ used for
cosets of $SU(2)
\times SU(2)$ with the (nonbold) square bracket notation used for the matrix commutator.)  From Eq.~(\ref{r54}) we see that the quotient of $SU(2)
\times
\bbox{(}SU(2)/B\bbox{)}$ by $\hat{S}/\tilde{B}$ is ${\Bbb R}P^3 \times
\bbox{(}SU(2)/B\bbox{)}$.  Applying Theorem~\ref{tm2} and recalling
Eq.~(\ref{r55}), we find

\begin{equation}
\Gamma 
= {SU(2) \times SU(2) \over \hat{S}}
={\bbox{(}SU(2) \times SU(2)\bbox{)}/\tilde{B} \over \hat{S}/\tilde{B}}
={SU(2) \times \bbox{(}SU(2)/B\bbox{)} \over \hat{S}/\tilde{B}}
= {\Bbb R}P^3 \times {\Bbb R}P^2.
\end{equation} 

\subsubsection{The class 2(ii) of planar symmetric tops}

For the class 2(ii), the matrices ${\sf A}$ and ${\sf B}$ are both in
$O(2)$ and satisfy $\det {\sf A} {\sf B} = 1$.  Thus, the isotropy
subgroup is

\begin{eqnarray}
S & = & \left \{
\begin{array}{c|c}
\left[
\begin{array}{c|c}
\;{\sf A}\; & \;{\sf 0}\; \\ \hline
{\sf 0} & {\sf B}
\end{array}
\right]
&
{\sf A}, {\sf B} \in O(2), \det {\sf A} {\sf B} = 1 
\end{array}
\right \} \nonumber \\
& = &
\left \{ {\sf I}, {\sf E}_1 \right \}
\left \{ 
\begin{array}{c|c}
\left[
\begin{array}{c|c}
\;{\sf A}\; & \;{\sf 0}\; \\ \hline
{\sf 0} & {\sf I}
\end{array}
\right]
& {\sf A} \in SO(2)
\end{array}
\right \}
\left \{ 
\begin{array}{c|c}
\left[
\begin{array}{c|c}
\;{\sf I}\; & \;{\sf 0}\; \\ \hline
{\sf 0} & {\sf B}
\end{array}
\right]
& {\sf B} \in SO(2)
\end{array}
\right \},
\label{r56}
\end{eqnarray} 
where we have factored $S$ into three factors.  The first two factors
multiply to give the group $S$ in Eq.~(\ref{r36}).  Thus, the double
cover of the first two factors is the group $\tilde{B}$ in
Eq.~(\ref{r52}).  Considering the last factor of Eq.~(\ref{r56}), we
note

\begin{eqnarray}
\pi\bbox{(}\exp(\theta {\sf \omega}_3), \exp( - \theta {\sf \omega}_3)\bbox{)}
& = & \exp\bbox{(} \theta({\sf J}_3 - {\sf L}_3)\bbox{)}
= \exp \left( 
2 \theta
\left[
\begin{array}{c|c}
\parbox{35pt}{\vskip 16pt \hskip 16pt ${\sf 0}$ \hskip 25pt \vskip 18pt } 
& {\sf 0} \\ \hline
{\sf 0} & 
\begin{array}{cc}
0 & -1 \\
1 & 0 
\end{array}
\end{array}
\right]
\right) \nonumber \\
& = & 
\left[
\begin{array}{c|c}
\parbox{35pt}{\vskip 16pt \hskip 16pt ${\sf I}$ \hskip 25pt \vskip 18pt } 
& {\sf 0} \\ \hline
{\sf 0} & 
\begin{array}{cc}
\cos 2\theta & -\sin 2\theta \\
\sin 2\theta  & \cos 2\theta 
\end{array}
\end{array}
\right],
\end{eqnarray}
where we used Eqs.~(\ref{r31}), (\ref{r34}), and (\ref{r27}).  Thus,
the double cover of the third factor in Eq.~(\ref{r56}) is the group

\begin{equation}
C
= \{ ({\sf U}, {\sf U}^\dagger)
| {\sf U} \in A \},
\end{equation}
where $A$ is the group defined in Eq.~(\ref{r76}).  Therefore, the
double cover of $S$ is given by the product of $\tilde{B}$ and $C$,

\begin{equation}
\hat{S}  
= \tilde{B} C
= \left \{ 
({\sf I}, {\sf I}),
({\sf \omega}_1, {\sf \omega}_1 ) 
\right \}
\tilde{A} C
= \left \{ 
({\sf I}, {\sf I}),
({\sf \omega}_1, {\sf \omega}_1 ) 
\right \}
D, 
\end{equation}
where we have used Eq.~(\ref{r52}) and where 

\begin{equation}
D = \tilde{A} C
= \{ ({\sf U}_1, {\sf U}_2)
| {\sf U}_1, {\sf U}_2 \in A \}
= U(1) \times U(1).
\end{equation}

We apply Theorem~\ref{tm2}, with $G = \hat{S}$, $H = D$, and $M =
SU(2) \times SU(2)$, to determine the topology of the kinematic orbit.
Using Eq.~(\ref{r40}), it is straightforward to verify that $D$ is
normal in $\hat{S}$, and hence $\hat{S}/D$ is a well-defined group
isomorphic to ${\Bbb Z}_2$.  Since $D = U(1) \times U(1)$, we
find

\begin{equation}
{SU(2) \times SU(2) \over D}
= {SU(2) \times SU(2) \over U(1) \times U(1)}
= {SU(2) \over A} \times {SU(2) \over A} 
= S^2 \times S^2,
\end{equation}
where we have used Eq.~(\ref{r64}).  From Eq.~(\ref{r65}), the action
of the nontrivial element $\bbox{[}{\sf
\omega}_1, {\sf \omega}_1\bbox{]} \in \hat{S}/D$ on $(\hat{\bf n}_1,
\hat{\bf n}_2) \in S^2 \times S^2$ is shown to be

\begin{equation}
\bbox{[}{\sf \omega}_1,{\sf \omega}_1\bbox{]}(\hat{\bf n}_1, \hat{\bf n}_2)
= (-\hat{\bf n}_1, -\hat{\bf n}_2).
\end{equation}
With this understanding of the action of ${\Bbb Z}_2$ on $S^2 \times
S^2$, we have

\begin{equation}
\Gamma = {SU(2) \times SU(2) \over \hat{S}}
= {\bbox{(}SU(2)\times SU(2)\bbox{)}/D \over \hat{S}/D}
= {S^2 \times S^2 \over {\Bbb Z}_2}.
\end{equation}

\subsubsection{The class 1 of collinear shapes}

In Sect.~\ref{s7} we showed that for collinear shapes $\Gamma =
SO(n-1)/O(n-2) = {\Bbb R}P^{n-2}$.  For $n=5$ this yields ${\Bbb
R}P^3$ and no more need be said.  However, for completeness and
analogy with the preceding cases, we show here how this result also
follows from Eq.~(\ref{r74}).

The matrix ${\sf B}$ can be any matrix in $O(3)$ and the matrix ${\sf
A}$ is the $1 \times 1$ matrix ${\sf A} = \det {\sf B}$.  As in the
analysis of class 2(i), we switch the positions of the blocks in ${\sf
K}$ containing ${\sf A}$ and ${\sf B}$.  Specifically, the isotropy
subgroup is

\begin{equation}
S = \left \{
\begin{array}{c|c}
\left[
\begin{array}{c|c}
\parbox{35pt}{\vskip 16pt \hskip 13pt ${\sf B}$ \hskip 25pt \vskip 18pt } 
& {\sf 0} \\ \hline
{\sf 0} & \det {\sf B}
\end{array}
\right]
&
{\sf B} \in O(3) 
\end{array}
\right \}
= \{ {\sf I}, -{\sf I} \}
\left \{
\begin{array}{c|c}
\left[
\begin{array}{c|c}
\parbox{35pt}{\vskip 16pt \hskip 13pt ${\sf B}$ \hskip 25pt \vskip 18pt } 
& {\sf 0} \\ \hline
{\sf 0} & \;1\;
\end{array}
\right]
&
{\sf B} \in SO(3) 
\end{array}
\right \}.
\end{equation}
Concerning the first factor, Eq.~(\ref{r49}) shows 

\begin{equation}
\pi(-{\sf I}, {\sf I}) = - {\sf I}.
\end{equation}
The second factor is the same as the group $S$ given in
Eq.~(\ref{r57}).  Thus, the double cover of the second factor is the
group $\tilde{E}$ given in Eq.~(\ref{r58}) and the double cover of $S$ is

\begin{equation}
\hat{S} 
= \{ ({\sf I}, {\sf I}), 
(-{\sf I}, {\sf I}) \}
\tilde{E}.
\end{equation} 

We apply Theorem~\ref{tm2} with $G = \hat{S}$, $H = \tilde{E}$, and $M
= SU(2) \times SU(2)$.  It is trivial to show that $\tilde{E}$ is
normal in $\hat{S}$ and hence $\hat{S}/\tilde{E}$ is a well-defined
group isomorphic to ${\Bbb Z}_2$.  Recall from Eq.~(\ref{r66}) that
$\bbox{(}SU(2) \times SU(2)\bbox{)}/\tilde{E} = SU(2)$.  The
nontrivial element $\bbox{[}-{\sf I}, {\sf I}\bbox{]} \in \hat{S}/\tilde{E}$ maps
$\bbox{[}{\sf U}_1, {\sf U}_2\bbox{]} \in \bbox{(}SU(2) \times
SU(2)\bbox{)}/\tilde{E}$ into $\bbox{[}-{\sf U}_1, {\sf U}_2\bbox{]}$.  Using the
diffeomorphism $f: \bbox{(}SU(2) \times SU(2)\bbox{)}/\tilde{E}
\rightarrow SU(2)$ of Eq.~(\ref{r53}), this results in the following
action on ${\sf U} \in SU(2)$,

\begin{equation}
\bbox{[}-{\sf I},{\sf I}\bbox{]}{\sf U} = - {\sf U}.
\end{equation}
Thus, we find

\begin{equation}
\Gamma = {SU(2) \times SU(2) \over \hat{S}}
= {\bbox{(}SU(2)\times SU(2)\bbox{)}/\tilde{E} \over \hat{S}/\tilde{E}}
= {SU(2) \over {\Bbb Z}_2}
= {\Bbb R}P^3.
\end{equation}

\section{Conclusions}

\label{s5}

For the general $n$-body problem, we have expressed a kinematic orbit
as the quotient of the kinematic group by the isotropy subgroup of
the shape in question.  We have computed these isotropy subgroups
explicitly.  For the three-, four-, and five-body cases, we have
represented the kinematic orbits in terms of simple well-studied
spaces of low dimension.  We have also showed that the kinematic orbit of a
collinear shape is ${\Bbb R}P^{n-2}$ for any $n$.

The natural next step for us to take is an analysis of body frame
singularities for $n\ge5$.  We envision such an analysis beginning, as
in the case of the three- and four-body
analysis\cite{Littlejohn98b,Littlejohn98a}, with a detailed study of
the principal axis frame and its singularities.  As in the previous
analysis, this would amount to finding the fundamental group of the
asymmetric top region of shape space and then relating the paths (or
more precisely the equivalence classes of paths) in this group to the
jumps in the principal axis frame.

In the three- and four-body problems, one can find a frame related to
the principal axis frame which has a smaller set of frame
singularities.  In particular, the frame jumps can be completely
eliminated.  A natural question is whether such a frame exists for $n
\ge 5$.  Extending this line of inquiry, another natural question is
which frames have the smallest set of singularities and what
constraints are placed on one's ability to move these singularities
around.  We believe that the study of frames restricted to the
kinematic orbits may shed some light on these issues.  For example, it
would be useful to know, in the language of fibre bundles, whether the
$SO(3)$ bundles defined over the kinematic orbits are trivial or not.

\section{Acknowledgements}
The authors gratefully acknowledge Professors Vincenzo Aquilanti and
Simona Cavalli for stimulating discussions motivating the present work
and for their kind hospitality during which this work was begun.  The
research in this paper was supported by the Engineering Research
Program of the Office of Basic Energy Sciences at the U. S. Department
of Energy under Contract No. DE-AC03-76SF00098.

\appendix

\section{Theorems on Lie Group Quotients}

\label{s6}

We present three theorems regarding the actions of Lie groups on
manifolds and the corresponding quotient spaces.  These results
provide a rigorous mathematical foundation for many of the steps
presented in the bulk of the paper.  The first result is a standard
theorem and is found, for example, in Bredon (Ref.~\cite{Bredon72},
p. 303, Corollary 1.3).

\begin{theorem}
\label{tm3}
Let $G$ be a compact Lie group acting smoothly on a smooth manifold
$M$.  Then the orbit through a point $x \in M$ is diffeomorphic to
$G/H$ where $H$ is the isotropy subgroup of $G$ at $x$.  (That is, $H$
contains all elements of $G$ which leave $x$ fixed.)
\end{theorem}

The next result is useful for simplifying the descriptions of several
manifolds appearing in the five-body problem.  It is similar to an
exercise of Bredon (Ref.~\cite{Bredon72}, p.~113, Exercise~9).

\begin{theorem}
\label{tm1}
Let $G$ be a compact Lie group and $H$ a Lie subgroup of $G$. Let
$\tilde{H}$ be the following Lie subgroup of $G \times G$,

\begin{equation}
\tilde{H} = \{ (h,h) | h \in H \}.
\label{r29}
\end{equation}
Of course, $\tilde{H}$ is trivially isomorphic to $H$.    Then, the smooth manifolds $(G \times
G)/\tilde{H}$ and $G \times (G/H)$ are diffeomorphic.
\end{theorem}

\noindent
\underline{Proof}

Assuming that $(G \times G)/\tilde{H}$ and $G/H$ are the right coset
spaces, we introduce the following notation for the right cosets

\begin{eqnarray}
\bbox{[}g_1, g_2\bbox{]}_{\tilde{H}} 
& = & \tilde{H}(g_1, g_2) \in (G \times G)/\tilde{H} 
\hskip 1cm g_1,g_2 \in G, \\ 
\; \bbox{[}g\bbox{]}_H & = & Hg \in G/H 
\hskip 1cm g \in G.
\end{eqnarray} 
We define a function $f: (G \times G)/\tilde{H} \rightarrow G \times (G/H)
$ acting on an arbitrary $\bbox{[}g_1,g_2\bbox{]}_{\tilde{H}} \in (G
\times G)/\tilde{H}$ by

\begin{equation}
f(\bbox{[}g_1,g_2\bbox{]}_{\tilde{H}}) = (g_2^{-1}g_1, \bbox{[}g_2\bbox{]}_H).
\label{r53}
\end{equation}
We assert that $f$ is a diffeomorphism.  First, we verify that $f$ is
well-defined on the coset space by noting

\begin{equation}
f(\bbox{[}hg_1,hg_2\bbox{]}_{\tilde{H}})
= \bbox{(}(h g_2)^{-1}(h g_1), \bbox{[}hg_2\bbox{]}_H\bbox{)}
= (g_2^{-1}g_1, \bbox{[}g_2\bbox{]}_H)
= f(\bbox{[}g_1,g_2\bbox{]}_{\tilde{H}}),
\end{equation}
where $h\in H$ is arbitrary.  Next, it is straightforward to verify
that the following function is well-defined and that it is the inverse of $f$

\begin{equation}
f^{-1}(g_1, \bbox{[}g_2\bbox{]}_H) = \bbox{[}g_2g_1,g_2\bbox{]}_{\tilde{H}},
\end{equation}
where $g_1,g_2 \in G$ are arbitrary.  Since both $f$ and $f^{-1}$ are
smooth, they are both diffeomorphisms.  ${\cal QED}$.

The following theorem is a refinement of an exercise in Bredon
(Ref.~\cite{Bredon72}, p. 67, Exercise 1) to the case of smooth
actions.  We omit the straightforward proof.

\begin{theorem}
\label{tm2}
Let $G$ be a compact Lie group and $H$ a normal Lie subgroup of $G$ so that
$G/H$ is itself a Lie group.  Let $G$ act smoothly upon a smooth
manifold $M$.  Assume that the isotropy subgroups of this action are
all conjugate to one another so that $M/G$ and $M/H$ are themselves
smooth manifolds.  Then, $G/H$ has a well-defined action on $M/H$
given by $\bbox{[}g\bbox{]}_H \bbox{[}x\bbox{]}_H = \bbox{[}gx\bbox{]}_H$, where $\bbox{[}g\bbox{]}_H \in G/H$ and $\bbox{[}x\bbox{]}_H \in
M/H$.  Furthermore, the following diffeomorphism holds

\begin{equation}
{M \over G} = {M/H \over G/H}.
\label{r60}
\end{equation}

\end{theorem}

\end{document}